\documentclass[prl,twocolumn,showpacs,nofootinbibs,superscriptaddress]{revtex4}
\usepackage{epsf}
\usepackage{amsmath}

\def\beq{\begin{equation}}
\def\eeq{\end{equation}}
\def\bea{\begin{eqnarray}}
\def\eea{\end{eqnarray}}
\def\beqa{\begin{equation}\begin{array}{l}}
\def\eeqa{\end{array}\end{equation}}

\def\Eqref#1{Eq.~(\ref{eq:#1})}

\def\half{\mbox{\small{$\frac{1}{2}$}}}

\def\barr{\left(\begin{array}{c}}
\def\earr{\end{array}\right)}
\def\bmat{\left(\begin{array}{cc}}
\def\emat{\end{array}\right)}

\def\ga{\gamma} 
 \def\De{\Delta}
\def\veps{\varepsilon}

\def\nn{\nonumber}

\def\mathscr{\mathcal}

\def\3d{3-D}

\def\OPE{1$\gamma $}
\def\TPE{2$\gamma $}

\begin{document}
\preprint{WM-05-118}
\preprint{JLAB-THY-05-}

\title{
Two-photon-exchange effects in the  electro-excitation of the $\Delta$ resonance}

\author{Vladimir Pascalutsa}
\email{vlad@jlab.org}
\affiliation{Physics Department, College of William and Mary,
Williamsburg, VA 23187, USA}
\affiliation{Theory Group, Thomas Jefferson National Accelerator Facility, 
Newport News, VA 23606, USA}
\author{Carl E.\ Carlson}
\email{carlson@physics.wm.edu}
\affiliation{Physics Department, College of William and Mary,
Williamsburg, VA 23187, USA}
\author{Marc Vanderhaeghen}
\email{marcvdh@jlab.org}
\affiliation{Physics Department, College of William and Mary,
Williamsburg, VA 23187, USA}
\affiliation{Theory Group, Thomas Jefferson National Accelerator Facility, 
Newport News, VA 23606, USA}

\date{\today}

\begin{abstract}
We evaluate the two-photon exchange contribution to the 
$e N \to e \Delta(1232) \to e \pi N$ process at large momentum transfer
with the aim of a precision study of the ratios of electric quadrupole (E2) and Coulomb quadrupole (C2) to 
the magnetic dipole (M1) $\gamma^* N \Delta$ transitions. 
We relate the two-photon exchange amplitude 
to the $N \to \Delta$ generalized parton distributions and obtain a 
quantitative estimate of the two-photon effects. 
The two-photon exchange corrections to the 
C2/M1 ratio depend strongly on whether
this quantity is obtained
from an interference cross section or from the Rosenbluth-type cross sections, 
in similarity with the elastic, $e N \to e N$, process.   
\end{abstract}


\maketitle
\thispagestyle{empty}

The properties of the  $\De(1232)$-resonance have been studied 
extensively in electron scattering off the nucleon. 
In all the up-to-date studies of the $\Delta$ excitation by electrons ($e N \to e \Delta$),
the electromagnetic interaction between the electron and the nucleon is assumed to be mediated by  
a single photon exchange (\OPE). In this Letter 
we study effects beyond this approximation, in particular the effects due to two-photon exchange (\TPE). 
The spectacular discrepancy between the polarization-transfer~\cite{Gayou02} 
and Rosenbluth-separation measurements~\cite{Arrington:2003df} of the nucleon form factors
is largely due to  \TPE-exchange effects~\cite{GV03,BMT03,YCC04,ourprd}.  Here we study
how such effects may alter the measurement of the electromagnetic $N\to \De$ transitions.  

The electromagnetic excitation of the $\Delta$ is dominated, at presently accessible $Q^2$, 
by a magnetic dipole ($M1$) transition, which in a simple quark model picture is described
by the spin flip of a quark in a zero orbital angular momentum state. 
The electric ($E2$) and Coulomb ($C2$) quadrupole $\ga N\De$ transitions
are weak at moderate $Q^2$ --- a few percent of the $M1$ strength.
The small $E2/M1$ and $C2/M1$ ratios are likely be prone to \TPE\ corrections which 
are roughly at the few percent level as well. Such corrections may be important to 
understand the perturbative QCD prediction which states that at large $Q^2$ 
the $E2/M1$ ratio approaches 100 percent~\cite{Carlson:1985mm} whereas the $C2/M1$ ratio 
is a constant corrected by double-logarithms of $Q^2$~\cite{Idilbi:2003wj}.

In the present work  
we consider the general formalism for the $e N \to e \Delta$ reaction beyond 
the \OPE-approximation and report the 
first model calculation of the \TPE\ effects 
in this process  using the partonic framework 
applied earlier for the elastic process~\cite{YCC04,ourprd}. As a result,
the \TPE\ amplitude is related to the $N \to \Delta$ generalized 
parton distributions (GPDs). 
\newline
\indent
When describing the $e N \to e \Delta$ process, 
the total number of helicity amplitudes is 32, which are reduced to 16 when 
using parity invariance. 
Furthermore, in a gauge theory lepton helicity is conserved to all orders in perturbation 
theory when the lepton mass is zero. We neglect the lepton mass, which   
reduces the number of helicity amplitudes to 8. 
In the \OPE\ approximation, there are three independent
helicity amplitudes which can be expressed in terms of 
three form factors of the $\ga^* N\De$ transition~\cite{Jones:1972ky,Caia:2004pm}:
$G_M^\ast (Q^2)$, $G_E^\ast (Q^2)$, and $G_C^\ast (Q^2)$, with $Q^2$ the 
photon virtuality.
Of special interest are the ratios of the electric ($G_E^*$)
and  Coulomb ($G_C^*$) quadrupole transitions to the
dominant magnetic dipole ($G_M^*$) transition:
\beq
R_{EM}  \equiv   - \frac{G^*_E}{G^*_M} , 
\quad \quad \quad R_{SM}  \equiv  - \frac{Q_+ \, Q_-}{(2 \, M_\Delta)^2} \, 
\frac{G^*_C}{G^*_M} ,
\label{eq:rsmdef} 
\eeq
where $Q_\pm = [Q^2 + (M_\Delta \pm M_N)^2]^{1/2}$, 
with nucleon mass $M_N$ = 0.938~GeV, 
and $\Delta$ mass $M_\Delta$ = 1.232~GeV.  

The $\gamma^* N \Delta$ transition form factors are  
usually studied in the pion electroproduction ($e N \to e \pi N$) process in the
$\De$-resonance region.
Denoting the invariant mass of the final $\pi N$ system by $W_{\pi N}$, 
we consider $W_{\pi N} = M_\Delta$. 
The 5-fold differential cross-section of this process is commonly written as: 
\begin{eqnarray}
\frac{d \sigma}{(d E_e^\prime \, d \Omega_e^\prime)^{lab} \, 
d \Omega_\pi} \,\equiv\, 
\Gamma_v \, \frac{d \sigma}{d \Omega_\pi},
\label{eq:crossv2}
\end{eqnarray}
where, in \OPE\ approximation, 
$d \sigma / d \Omega_\pi$ has the interpretation of a 
$\gamma^* N \to \pi N$ virtual photon absorption cross section,
and the virtual photon flux factor $\Gamma_v$ is given by:
\begin{eqnarray}
\Gamma_v = \frac{e^2}{(2 \pi)^3} \, 
\left(\frac{E_e^\prime}{E_e}\right)^{lab} \, 
\frac{(W_{\pi N}^2 - M_N^2) / (2 M_N)}{Q^2 \, (1 - \varepsilon)},
\label{eq:crossv3} 
\end{eqnarray}
where 
$E_e^{lab}$ ($E_e^{\prime \, lab}$) are the initial (final) electron {\it lab} 
energies, $e$ is the electric charge, 
and $\varepsilon$ denotes the photon polarization parameter. 
 The pion angles $(\theta_\pi, \Phi)$ are defined in the $\pi N$ {\it c.m.} 
frame, with $\theta_\pi$ the pion polar angle and $\Phi$  
the angle between the hadron and lepton planes ($\Phi = 0$  
corresponds with the pion emitted in the same half-plane as the leptons).   
For unpolarized nucleons, the cross section at $W_{\pi N} = M_\Delta$ can, 
in general, be parametrized as~:
\begin{eqnarray}
\frac{d \sigma}{d \Omega_\pi}&=& \sigma_{0}    
+ \varepsilon \, \cos (2 \Phi) \, \sigma_{TT} 
+\, (2 h) \, \varepsilon \, \sin (2 \Phi) \, \sigma_{TT i} \nonumber \\
&+&\sqrt{2 \varepsilon}\,\veps_+ \cos \Phi \, \sigma_{LT}
+ (2 h)  \sqrt{2 \varepsilon }\, \veps_- \sin \Phi \, \sigma_{LT i} , \ \ 
\label{eq:crossv6}
\end{eqnarray}
with $h = \pm 1/2$ the lepton helicity and 
$\varepsilon_\pm \equiv  \sqrt{1 \,\pm\, \varepsilon}$.
The cross sections $\sigma_{TT i}$ and $\sigma_{LT i}$ 
are new responses which appear beyond the \OPE-exchange approximation.
For the cross-sections $\sigma_0$, $\sigma_{TT}$, $\sigma_{LT}$
the \TPE\ exchange induces corrections of order $e^2$ relative
to \OPE. 

\begin{figure}[t]
\centerline{  \epsfxsize=6cm
  \epsffile{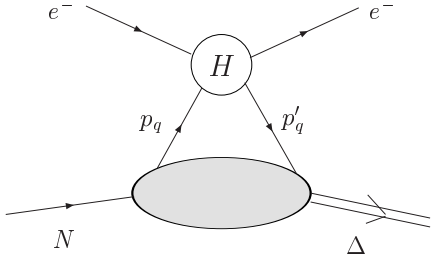} 
}
\caption{Handbag approximation for the $e N \to e \Delta$ process  
 at large momentum transfers. In the partonic scattering process 
($H$), the lepton scatters from quarks with momenta $p_q$ and $p'_q$. 
The lower blob represents the $N \to \Delta$ GPD's.}
\label{fig:handbag}
\end{figure}

In order to evaluate the \TPE\ contribution 
to the $N \to \Delta$ electroproduction amplitudes 
at large momentum transfers, we will consider a partonic model 
in the handbag approximation, illustrated in Fig.~\ref{fig:handbag}, 
as studied before for the $e N \to e N$ process~\cite{YCC04,ourprd}. 
This calculation  involves a hard scattering 
subprocess on a quark, 
which is then embedded in the proton by means of
the $N \to \Delta$ GPDs~\cite{Fra00,GPV01,GMV03}. 
We consider only the vector GPD $H_M^{(3)}$ and 
axial-vector GPD $C_1^{(3)}$ 
(the subscript 3 denotes their isovector character) 
since they  dominate in the large $N_c$ limit. The handbag amplitude can then be specified in
terms of the following two characteristic integrals~:
\begin{eqnarray}
A^* &=& \int_{-1}^1 \frac{dx}{x}  \    
	\left[ \frac{\hat s - \hat u}{Q^2} g_M^{\,hard} + g_A^{(2\gamma)}\right]
	\sqrt{\frac{2}{3}}   \,\frac{1}{6} \, H_M^{(3)}	\,, 
			 \\
C^* &=& \int_{-1}^1 \frac{dx}{x} \ \left[ \frac{\hat s - \hat u}{Q^2} g_A^{(2\gamma)} + g_M^{\,hard}\right]
	\, \mathrm{sgm}(x) \,\frac{1}{6}\,  C_1^{(3)} ,	\ \ \  
\label{eq:ac}
\end{eqnarray}
where the factor 1/6 results from the quadratic
quark charge combination \( (e_{u}^{2}-e_{d}^{2})/2 \), 
and all hard scattering quantities in the square brackets are given in 
Ref.~\cite{ourprd}. 

The GPDs \( H_{M}^{(3)} \) ($C_{1}^{(3)}$) are linked  
with the \( N\rightarrow \Delta  \) vector (axial-vector) 
transition form factors \( G_{M}^{*} \) ($C_{5}^{A}$, 
as introduced by Adler~\cite{Adl75}) respectively through the sum rules~:
\begin{eqnarray}
\int _{-1}^{1} dx\, H_{M}^{(3)}(x,0 ,Q^2)
&=& 2\, G_{M}^{*}(Q^2) ,
\label{eq:vec-sumrule} 
\\
\int _{-1}^{1} dx\, C_{1}^{(3)}(x,0 ,Q^2)\,  
& = & 2\, C_{5}^{A}(Q^2)\, .
\label{eq:axial-sumrule} 
\end{eqnarray}
At \( Q^2=0 \), $G_M^*$ is extracted
from pion photoproduction experiments as~: \( G_{M}^{*}(0)\simeq 3.02 \)
\cite{Tia00}.
For small \( Q^2 \), PCAC leads to a dominance
of the form factors \( C_{5}^{A} \), for which 
a Goldberger-Treiman relation for the \( N\rightarrow \Delta  \)
transition yields~: 
$ \sqrt{3/2} \, C_{5}^{A}(0)= g_{A}\,f_{\pi N\Delta }/(2f_{\pi NN})$.
 Using the phenomenological values \( f_{\pi N\Delta }\simeq 1.95 \),
\( f_{\pi NN} \simeq 1.00 \), and \( g_{A}\simeq 1.267 \) one obtains \( C_{5}^{A}(0)\simeq 1.01 \).

We can now compute the \TPE\ effects in observables.  
We start by multipole expanding Eq.~(\ref{eq:crossv6}) 
for the $e p \to e \Delta^+ \to e \pi N$ process as~:
\begin{eqnarray}
\sigma_0
&=& A_0 \,    
+ \half (3 \cos^2 \theta_\pi - 1) \, A_2 ,
 \nn \\
\sigma_{TT}
&=& \sin^2 \theta_\pi \, C_0 , \nn\\
\sigma_{LT}
&=& \half  \sin(2 \theta_\pi) \, D_1,
\label{eq:crossv7} \\
\sigma_{TT i}
&=& \sin^2 \theta_\pi \, C_{0 i} , \nn\\
\sigma_{LT i}&
= &\half   \sin(2 \theta_\pi) \, D_{1 i},
\nonumber
\end{eqnarray}
where $A_0$ can 
be written as: 
\begin{eqnarray}
A_{0 } = {\cal I} \, \frac{e^2}{4 \pi} \frac{Q_-^2}{4 \, M_N^2} 
\frac{(M_\Delta + M_N)}{(M_\Delta - M_N)} \frac{1}{M_\Delta \, \Gamma_\Delta} 
(G_M^*)^2\, \sigma_R,
\label{eq:aom}
\end{eqnarray} 
where $\Gamma_\Delta \simeq 0.120$~GeV is the $\Delta$ width, and  
${\cal I}$ denotes an isospin factor which depends on the 
final state in the $\Delta^+ \to \pi N$ decay as~: 
${\cal I}(\pi^0 p) = 2/3$ and ${\cal I}(\pi^+ n) = 1/3$.
Furthermore in \Eqref{aom}, the reduced cross section $\sigma_R$,  
including \TPE\ corrections evaluated in the handbag model,  
is given by~:
\begin{eqnarray}
&&\hspace{-0.3cm}\sigma_R = 1 + 3 \, R_{EM}^2 + \varepsilon \, 
\frac{16 \, M_\Delta^2 \, Q^2}{Q_+^2 \, Q_-^2} \, R_{SM}^2 
\label{eq:crossred} \\
&&\hspace{-0.3cm}+\, \frac{1}{G_M^*}  \sqrt{\frac{2}{3}}  
\left[ \frac{A^*}{2}   
	\frac{Q^2 \varepsilon_+ \varepsilon_- }{Q_+ \, Q_-}  
+ 2 \, C^*  \frac{Q^2}{Q_-^2}  \varepsilon_-^2 \, 
\frac{M_N}{M_N + M_\Delta} \right]. 
\nonumber 
\end{eqnarray}

We next discuss the $2 \gamma$ corrections to $R_{EM}$ and $R_{SM}$ 
as extracted from $\sigma_{TT}$ and $\sigma_{LT}$. 
Experimentally, these ratios have been extracted at $W_{\pi N} = M_\Delta$ using~:
\begin{eqnarray}
R_{EM}^{exp,I} &=& \frac{3 A_2 - 2 C_0}{12 A_0}
	\stackrel{1\gamma}{=}  
	R_{EM}  + \varepsilon \, \frac{4 M_\Delta^2 Q^2}{Q_+^2\, Q_-^2} R_{SM}^2
	+ \ldots
	\label{eq:rem1a} \nonumber \\
R_{SM}^{exp} &=& \frac{Q_+ \, Q_-}{Q \, M_\Delta} \frac{D_1}{6 A_0}
	\stackrel{1\gamma}{=} R_{SM} - R_{SM} R_{EM} + \ldots 
\label{eq:rsm1}
\end{eqnarray}

\noindent where the omitted terms involve cubic products of $R_{EM}$ and $R_{SM}$.  
These formulas are usually applied~\cite{Joo:2001tw} 
by neglecting the smaller quantities 
$R_{SM}^2$ and $R_{EM} \cdot R_{SM}$. 
We will keep the quadratic terms here and show that the 
$R_{SM}^2$ term leads to a non-negligible contribution 
in the extraction of $R_{EM}$.  
A second way of extracting $R_{EM}$, which avoids corrections at the one-photon level, is~: 
\begin{eqnarray}
R_{EM}^{exp,II} = \frac{-(A_0 - A_2) - 2 \, C_0}{ 3(A_0 - A_2) - 2 \, C_0 }
&\stackrel{1\ga}{=}  R_{EM} . 
\label{eq:rem1b} 
\end{eqnarray}

We can usefully denote the corrections to $R_{EM}$ and $R_{SM}$ generically by~:
\begin{eqnarray}
R \,&\simeq&\, R^{exp} + \delta R^{1 \gamma} 
+ \delta R^{2 \gamma} .
\end{eqnarray}

\noindent  The term $\delta R^{1 \gamma}$ 
denotes the corrections due to the quadratic terms 
in Eqs.~(\ref{eq:rem1a},\ref{eq:rem1b}), which are~: 
\begin{eqnarray}
\delta R_{EM}^{1 \gamma, I} &=& 
- \varepsilon \, \frac{4 \, M_\Delta^2 \, Q^2}{Q_+^2 \, Q_-^2} \,
R_{SM}^2 ,
\nonumber \\
\delta R_{EM}^{1 \gamma, II} &=& 0 , \\
\delta R_{SM}^{1 \gamma} &=& R_{EM} \cdot R_{SM} .\nn
\end{eqnarray}

\noindent The two-photon exchange corrections to $R_{EM}$ 
and $R_{SM}$ are obtained in the handbag model and are~:
\begin{eqnarray}
\delta R_{EM}^{2 \gamma, I} &=& - 
\frac{1}{8} \sqrt{\frac{3}{2}} \frac{Q^2}{Q_+ \, Q_-} \, 
\frac{\varepsilon_-^3 \, \varepsilon_+}{\varepsilon} \, 
\frac{1}{G_M^*} \, A^*  \nonumber \\
&+& \frac{1}{4} \sqrt{\frac{2}{3}} \, \frac{Q^2}{Q_-^2} \, 
\frac{\varepsilon_-^2 \, \varepsilon_+^2}{\varepsilon} \, 
\frac{M_N}{(M_N + M_\Delta)} \, \frac{1}{G_M^*} \, C^* \, ,
\nonumber \\
\delta R_{EM}^{2 \gamma, II} &=& 2 \,
\delta R_{EM}^{2 \gamma, I} ,
\nonumber \\
\delta R_{SM}^{2 \gamma} &=& 
\sqrt{\frac{2}{3}} \, \frac{(Q^2 - M_\Delta^2 + M_N^2)}{4 \, M_\Delta^2} \, 
\frac{Q_+}{Q_-} \, \frac{1}{\sqrt{2 \, \varepsilon}}
\frac{\varepsilon_-^2}{\varepsilon_+} \, \nonumber \\
&\times& \frac{M_N}{(M_N + M_\Delta)} 
\, \frac{1}{G_M^*} \, C^* \, .
\end{eqnarray}

\begin{figure}[t]
\centerline{  \epsfxsize=7cm
  \epsffile{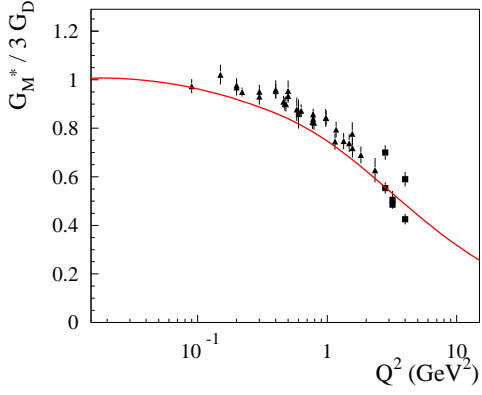} 
}
\caption{\label{fig:gmdelmregge}(Color online) 
$Q^2$ dependence of the form factor $G_M^*$ divided by $3 \cdot G_D$ with 
$G_D = 1/(1 + Q^2/0.71)^2$. The calculation corresponds with the modified 
Regge model ($\alpha'_2 = 1.3$ GeV$^{-2}$).  
The data for $G_M^{*}$ are from the compilation of Ref.~\cite{Tia00}. 
}
\end{figure}

To provide numerical estimates for the $2 \gamma$ corrections, 
we need a model for the two `large' GPDs
which appear in the integrals $A^*$ and $C^*$.   
Here we will be guided by the large
\( N_{c} \) relations discussed in~\cite{Fra00,GPV01}. These relations
connect the \( N\rightarrow \Delta  \) GPDs \( H_{M}^{(3)} \) 
and \( C_{1}^{(3)} \) 
to the \( N\rightarrow N \) isovector GPDs \( E^{u}-E^{d} \),
and \( \tilde{H}^{u}-\tilde{H}^{d} \) respectively as~:
\begin{eqnarray}
H_{M}^{(3)}(x, 0 , Q^2) & = & 2  \frac{G_M^*(0)}{\kappa_V}  
 \left[ E^{u} - E^{d}\right](x, 0 ,Q^2) ,
\nonumber \\
C_{1}^{(3)}(x, 0 , Q^2) & = & 
\sqrt{3}\left[ \tilde{H}^{u} - \tilde{H}^{d}\right](x, 0 , Q^2) ,
\label{Eq_DEL.8} 
\end{eqnarray}
with $\kappa_V = \kappa_p - \kappa_n = 3.706$.
For the nucleon GPDs $E^q$ and $\tilde H^q$ appearing in 
Eq.~(\ref{Eq_DEL.8}), 
we use the modified Regge model~\cite{guidal}, which was applied   
before~\cite{ourprd} to estimate $2 \gamma$ 
corrections to the $e N \to e N$ process. 
As an example, we show in Fig.~\ref{fig:gmdelmregge} 
the form factor $G_M^*$, obtained 
by evaluating the sum rule of Eq.~(\ref{eq:vec-sumrule}) using  
the modified Regge model for the GPD $H_M^{(3)}$, adjusting the Regge 
slope parameter as $\alpha'_2 = 1.3$ GeV$^{-2}$. 
One sees that a good description is obtained over the whole range of $Q^2$. 

\begin{figure}[t]
\centerline{ \epsfxsize=8.5cm  \epsffile{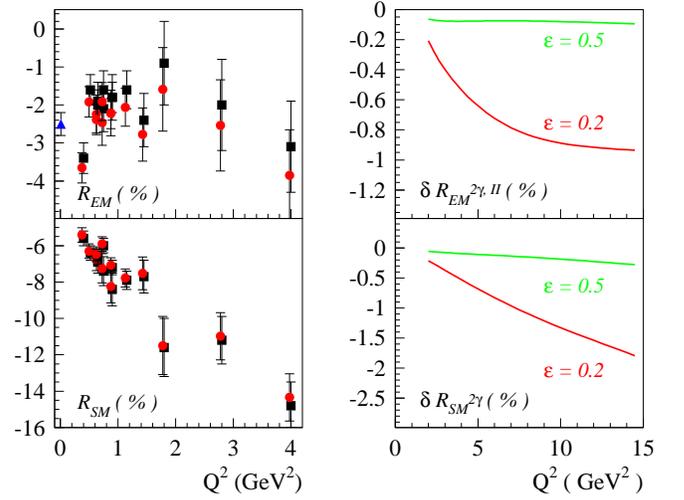} 
}
\caption{\label{fig:remrsmq}(Color online)  
Left panels:  $Q^2$ dependence of the \OPE\ corrections, $\delta R^{1 \gamma}$, to 
$R_{EM}$ (upper panel) and $R_{SM}$ (lower panel). 
The squares are the uncorrected data from~\cite{Joo:2001tw} 
and (two highest $Q^2$ points)~\cite{Frolov:1998pw}. 
The error bars reflect the statistical error only.   
The red circles include the corrections 
$\delta R_{EM}^{1 \gamma, I}$ and $\delta R_{SM}^{1 \gamma}$. 
The triangle is the real photon point from~\cite{Mainz97}.
Right panels: $Q^2$ dependence of the \TPE\ corrections to 
$R_{EM}$ (upper panel) and $R_{SM}$ (lower panel) at 
different values of $\varepsilon$ as indicated.
}
\end{figure}

In Fig.~\ref{fig:remrsmq}, we show the effect of the \OPE\ and \TPE\ 
corrections on $R_{EM}$ and $R_{SM}$. For the \OPE\ correction, we see that 
the effect on $R_{SM}$ is negligible, whereas it yields a systematic 
downward shift of the $R_{EM}$ result. This shift becomes more pronounced 
at larger $\varepsilon$, and it is found that for 
$Q^2$ around 5 GeV$^2$ such as for the upcoming data of Ref.~\cite{Ungaro}, 
it decreases $R_{EM}$ by around 1 \%. To avoid such a correction, it calls 
for extracting $R_{EM}$ according to the procedure $II$ as we outlined above. 
Furthermore, we show the \TPE \ corrections 
in Fig.~\ref{fig:remrsmq} (right panels), 
estimated using the modified Regge GPD model. 
We see that the \TPE \ effects are mainly pronounced at small $\varepsilon$ 
and larger $Q^2$. For $R_{EM}$ they are well below 1 \%, whereas they yield 
a negative correction to $R_{SM}$ by around 1 \%, when $R_{SM}$ is extracted 
from $\sigma_{LT}$ according to Eq.~(\ref{eq:rsm1}). 

\begin{figure}[t]
\centerline{  \epsfxsize=9cm
  \epsffile{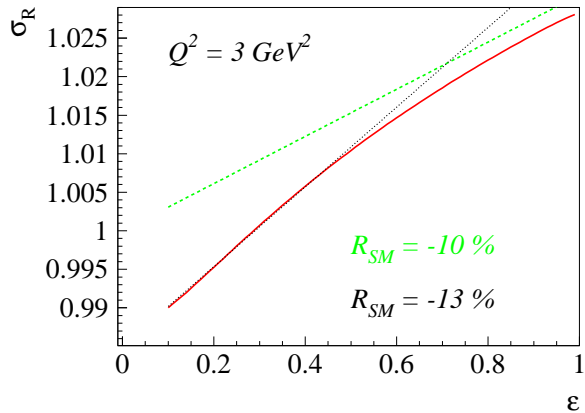} 
}
\vspace{-0.5cm}
\caption{\label{fig:remrsm3}(Color online)   
Rosenbluth plot of the reduced cross section $\sigma_R$ 
of Eq.~(\ref{eq:crossred}) for the $e N \to e \Delta$ 
reaction at $Q^2 = 3$~GeV$^2$.
The dashed curve corresponding with $R_{SM} = -10 \%$  
is the \OPE \ result, 
whereas the solid curve represents the result including 
\TPE \ corrections for the same values of $R_{SM}$. 
The dotted curve corresponding with  $R_{SM} = -13 \%$ 
corresponds with a linear fit to the total result in an intermediate 
$\varepsilon$ range.  
}
\end{figure}

Another way to obtain $R_{SM}$ is from a Rosenbluth-like analysis of
the cross section $\sigma_0$ for the $e N \to e \Delta$ reaction.  In
Fig.~\ref{fig:remrsm3}, we show the reduced cross section from
Eq.~(\ref{eq:crossred}). One sees how a different extraction of
$R_{SM}$ can yield a significantly different result.  Starting from a
value of $R_{SM}$ (e.g., --10 \%) extracted from $\sigma_{LT}$, adding
in the two-photon exchange corrections gives a sizable change in the
slope of the Rosenbluth plot.  When fitting the total result by a
straight line in an intermediate $\varepsilon$ range, one extracts a
value of $R_{SM}$ around  3 percentage units lower than  its value as
extracted from $\sigma_{LT}$. This is sizable, as it corresponds with
a 30\% correction on the absolute value of $R_{SM}$. The situation is
similar to extracting the elastic proton form factor ratio $G_E/G_M$
using the Rosenbluth method~\cite{GV03}. It will be interesting to
confront this to new Rosenbluth separation data in the $\Delta$ region
up to $Q^2 \simeq 5$~GeV$^2$ which are presently under
analysis~\cite{Tvaskis}. 

Summarizing, in this work we estimated 
the \TPE \ contribution to the 
$e N \to e \Delta(1232) \to e \pi N$ process at large momentum transfer
in a partonic model. 
We related the \TPE \ amplitude 
to the $N \to \Delta$ generalized parton distributions. 
For $R_{EM}$, the \TPE \ corrections were found to be small, 
below the 1 \% level. We showed however that the neglect of a quadratic term  
$R_{SM}^2$ in the usual method to extract $R_{EM}$ 
yields corrections at the 1 \% level.    
The \TPE \ corrections to $R_{SM}$ were found to be substantially different 
when extracting this quantity 
from an interference cross section or from Rosenbluth type cross sections, 
as has been observed before for the elastic $e N \to e N$ process.   
It will be interesting to confront these results with 
upcoming new Rosenbluth separation data at intermediate $Q^2$ values 
in order to arrive at a precision extraction of the large $Q^2$  
behavior of the $R_{EM}$ and $R_{SM}$ ratios.

\begin{acknowledgments}
The authors thank A.~Afanasev for useful discussions. 
This work is supported in part by DOE grant no.\
DE-FG02-04ER41302, contract DE-AC05-84ER-40150 under
which SURA operates the Jefferson Laboratory, and
by the National Science Foundation under grant PHY-0245056 (C.E.C.).  
\end{acknowledgments}


\begin{thebibliography}{99}




\bibitem{Gayou02}
O.~Gayou {\it et al.}  [Jefferson Lab Hall A Collaboration],
Phys.\ Rev.\ Lett.\  {\bf 88}, 092301 (2002).


\bibitem{Arrington:2003df}
  J.~Arrington,
  Phys.\ Rev.\ C {\bf 68}, 034325 (2003).


\bibitem{GV03}
P.~A.~M.~Guichon and M.~Vanderhaeghen,
Phys.\ Rev.\ Lett.\  {\bf 91}, 142303 (2003).


\bibitem{BMT03}
P.~G.~Blunden, W.~Melnitchouk and J.~A.~Tjon,
Phys.\ Rev.\ Lett.\  {\bf 91}, 142304 (2003).


\bibitem{YCC04}
Y.~C.~Chen, A.~Afanasev, S.~J.~Brodsky, C.~E.~Carlson and M.~Vanderhaeghen,
Phys.\ Rev.\ Lett.\  {\bf 93}, 122301 (2004).


\bibitem{ourprd}
 A.~V.~Afanasev, S.~J.~Brodsky, C.~E.~Carlson, Y.~C.~Chen and M.~Vanderhaeghen,
  Phys.\ Rev.\ D {\bf 72}, 013008 (2005).

\bibitem{Joo:2001tw}
  K.~Joo {\it et al.}  [CLAS Collaboration],
  Phys.\ Rev.\ Lett.\  {\bf 88}, 122001 (2002).

\bibitem{Carlson:1985mm}
  C.~E.~Carlson,
  Phys.\ Rev.\ D {\bf 34}, 2704 (1986).

\bibitem{Idilbi:2003wj}
  A.~Idilbi, X.~d.~Ji and J.~P.~Ma,
  Phys.\ Rev.\ D {\bf 69}, 014006 (2004).

\bibitem{Jones:1972ky}
H.~F.~Jones and M.~D.~Scadron,
Ann.\ Phys.\  {\bf 81}, 1 (1973).

  \bibitem{Caia:2004pm}
G.~L.~Caia, V.~Pascalutsa, J.~A.~Tjon and L.~E.~Wright,
  Phys.\ Rev.\ C {\bf 70}, 032201 (2004).

\bibitem{GPV01} 
K. Goeke, M.V. Polyakov and M. Vanderhaeghen, 
Prog. Part. Nucl. Phys. \textbf{47,} 401 (2001). 

\bibitem{Fra00} 
L.~L. Frankfurt, M.~V. Polyakov, M. Strikman, and M. Vanderhaeghen, 
Phys. Rev. Lett. \textbf{84}, 2589 (2000). 

\bibitem{GMV03}
  P.~A.~M.~Guichon, L.~Mosse and M.~Vanderhaeghen,
  Phys.\ Rev.\ D {\bf 68}, 034018 (2003).

\bibitem{Tia00}
L.~Tiator, D.~Drechsel, O.~Hanstein, S.~S.~Kamalov and S.~N.~Yang,
Nucl.\ Phys.\ A {\bf 689}, 205 (2001).


\bibitem{Frolov:1998pw}
V.~V.~Frolov {\it et al.},
Phys.\ Rev.\ Lett.\  {\bf 82}, 45 (1999).

\bibitem{Adl75} 
S. Adler, Ann. Phys. (N.Y.) \textbf{50}, 189 (1968). 


\bibitem{guidal} 
M. Guidal, M. Polyakov, A. Radyushkin, and M. Vanderhaeghen, 
arXiv:hep-ph/0410251.

\bibitem{Mainz97} 
R.~Beck {\em et al.},  Phys.\ Rev.\ Lett.\  {\bf 78},
606 (1997); Phys.\ Rev.\ C  {\bf 61}, 035204 (2000).

\bibitem{Ungaro}
M. Ungaro {\it et al.} [CLAS Collaboration], to be published. 

\bibitem{Tvaskis}
V. Tvaskis, talk presented at the JLab/INT Workshop 
``Precision Electroweak Physics'', College of William and Mary, 
August 15-17, 2005. 

\end{thebibliography}
\end{document}